\documentclass[
aps,%
final,%
notitlepage,%
oneside,%
onecolumn,%
nobibnotes,%
nofootinbib,%
superscriptaddress,%
noshowpacs,%
centertags]%
{revtex4}%
\usepackage{graphicx}
\usepackage{amsmath}
\usepackage{amssymb}

\begin{document}

\title{Effect of weak magnetic field on polariton-electron scattering
in semiconductor microcavities}%

\author{\firstname{V.V.}~\surname{Bilykh}}
\email{belykh@sci.lebedev.ru}
\affiliation{%
P.N. Lebedev Physical Institute - 119991 Moscow, Russia
}%

\begin{abstract}
We theoretically calculate the polariton linewidth associated with
the polariton-electron scattering in a microcavity in presence of
a magnetic field perpendicular to the microcavity plane. It is
shown that the polariton linewidth oscillates as a function of the
magnetic field magnitude and the polariton-electron scattering
rate can be not only decreased but also increased by the magnetic
field. The possible applications of such an effect are discussed.
\end{abstract}

\maketitle

\section{Introduction}

Microcavity (MC) polaritons are quasiparticles resulting from the
strong exciton-photon coupling. They have attracted considerable
interest since the observation of the strong coupling regime in
semiconductor MC \cite{Weisbuch}. MC polaritons have bosonic
nature and are characterized by the steep dispersion for small
wave vectors (Fig.\ref{dispersion}a). These properties predict the
possibility of macrofilling of states at the bottom of the lower
polariton (LP) branch (MC polariton bose condensation) and
creation of ``polariton laser''. But it was not achieved under
nonresonant optical excitation of MC (above the band gap of the
semiconductor) due to the small polariton lifetime near the bottom
of the LP branch as compared to the relaxation time \cite{Butte}.
Under resonant excitation near the inflection point of the LP
branch, the macrofilling of states at the bottom of the LP branch
was achieved as a result of the four-wave-mixing (FWM) effect
\cite{Stevenson} which is a stimulated scattering of the pump
polaritons into the ``signal'' and ``idler'' modes due to the
polariton-polariton interaction. The experiments
\cite{Krizhanovskii,Krizhanovskii1} show that nonstimulated
polariton scattering processes govern the FWM threshold and its
efficiency. For small values of the LP branch depth
(Fig.\ref{dispersion}a), when the FWM is most efficient,
nonstimulated scattering increase the FWM threshold and decrease
its efficiency \cite{Krizhanovskii}. It was shown theoretically
\cite{Malpuech} and experimentally \cite{Lagoudakis,Tartakovskii}
that one of the most efficient polariton scattering mechanisms is
scattering by free electrons present in a quantum well (QW). It is
of obvious interest on the one hand to suppress the
polariton-electron scattering in FWM experiments and on the other
hand to enhance its efficiency for nonresonant system excitation.
We show that the polariton-electron scattering rate can be
decreased as well as increased by a magnetic field, applied
perpendicular to the MC plane. The obtained oscillatory dependence
of the polariton-electron scattering rate on the magnetic field
magnitude can also be used for distinguishing the
polariton-electron scattering from other possible polariton
scattering mechanisms.

The polariton-electron scattering can be characterized by a
polariton linewidth due to the electron scattering. First the
polariton linewidth is calculated without a magnetic field, then
the effect of a weak magnetic field (magnetic length is much
larger than exciton Bohr radius) is taken into account.

\section{Polariton-electron scattering without magnetic field}

\begin{figure}
\includegraphics{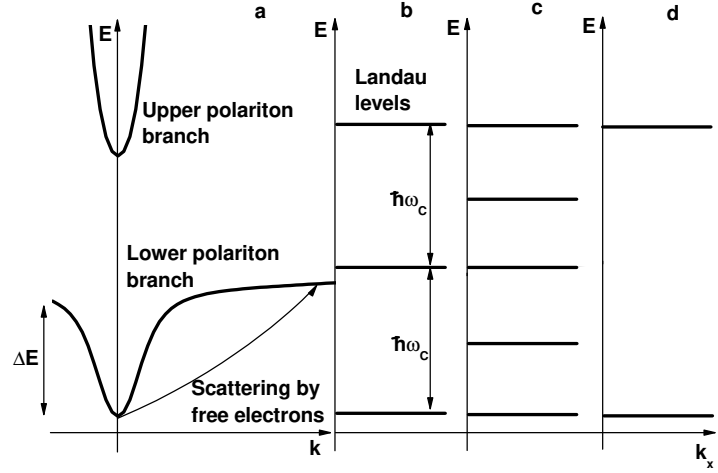}%
\caption{a) MC polariton dispersion curves; b),c),d) position of
free QW electron Landau levels for the different magnitudes of the
magnetic field. $\Delta E$ is the LP branch depth (separation
between the free exciton level and the
minima of the LP branch).}%
\label{dispersion}
\end{figure}
We calculate the MC polariton linewidth near the bottom of the LP
branch due to the free QW electron scattering, neglecting the
influence of the upper polariton branch states and localized
exciton states. The spin distribution of particles is assumed to
be isotropic. Polariton-electron scattering results from Coulomb
interaction of an electron with an exciton which is a component of
a polariton. For high enough temperature of the electron gas ($T
\sim \Delta E$), polaritons from the region $k \approx 0$ scatter
into the reservoir of the excitonlike states with high $k$
(Fig.\ref{dispersion}a). The probability for the polariton to
leave the state with the wave vector $\vec{K}_i$ is calculated
using the Fermi Golden rule:
\begin{equation}
\label{linew} %
\gamma_{\vec{\textbf{K}}_i}= 4 \pi \sum_{\vec{\textbf{k}}_i,\vec{\textbf{k}}_f,s}%
p_s \cdot %
|M_{\vec{\textbf{K}}_i,\vec{\textbf{k}}_i,\vec{\textbf{Q}},s}|^2
\cdot F_{\epsilon_i} \cdot (1-F_{\epsilon_f}) \cdot
\delta(E_f+\epsilon_f-E_i-\epsilon_i),
\end{equation}
where
$\vec{\textbf{Q}}=\vec{\textbf{K}}_f-\vec{\textbf{K}}_i=\vec{\textbf{k}}_i-\vec{\textbf{k}}_f$;
$\vec{\textbf{K}}, \vec{\textbf{k}}$ are the wave vectors and $E,
\epsilon$ are the energies of the polariton and the electron
respectively; index ``$i$'' corresponds to the initial state,
``$f$'' corresponds to the final state; $s=\pm$ is the spin
configuration of a free electron and an electron which is a
component of an exciton: sign ``$+$'' corresponds to the triplet
configuration, sign ``$-$'' corresponds to the singlet
configuration; $p_s$ is the probability for the system to have the
spin configuration $s$, for the isotropic spin distribution of
particles $p_+=\frac{3}{4}, p_-=\frac{1}{4}$ \cite{Landau};
$F_\epsilon=1/(e^{\frac{\epsilon-\mu}{T}}+1)$ is the Fermi-Dirac
electron distribution function;
$M_{\vec{\textbf{K}}_i,\vec{\textbf{k}}_i,\vec{\textbf{Q}},s}$ is
the matrix element of the Coulomb potential of the
polariton-electron interaction.

We use the following wave functions (WF) to calculate the matrix
element
$M_{\vec{\textbf{K}}_i,\vec{\textbf{k}}_i,\vec{\textbf{Q}},s}$:
\begin{equation}
\label{WF} |\vec{\textbf{K}},\vec{\textbf{k}}\rangle=(C_K
\hat{a}_{\vec{\textbf{K}}}^+ + X_K
\hat{b}_{\vec{\textbf{K}}}^+)\cdot \hat{c}_{\vec{\textbf{k}}}^+ |0
\rangle,
\end{equation}
where $\hat{a}^+,\hat{b}^+,\hat{c}^+$ are the creation operators
of a photon, an exciton, and an electron respectively; $C_K, X_K$
are the Hopfield coefficients of a photon and an exciton
respectively \cite{Hopfield, Sermage}. In Eq. (\ref{WF}), in
brackets, is the polariton creation operator. The exciton creation
operator is given by the following expression (we note that the
exciton is in its ground state):
\begin{equation}
\hat{b}_{\vec{\textbf{K}}}^+=\sum_{\vec{\textbf{k}}_0}\phi_{\alpha
\vec{\textbf{K}}-\vec{\textbf{k}}_0} \cdot
\hat{c}_{\vec{\textbf{k}}_0}^+
\hat{d}_{\vec{\textbf{K}}-\vec{\textbf{k}}_0}^+ ,
\end{equation}
where $\hat{d}^+$ is the hole creation operator; $\alpha$ is the
ratio of electron and exciton effective masses;
$\phi_{\vec{\textbf{k}}}=\sqrt{\frac{8 \pi \lambda^2}{S}} \cdot
(1+(\lambda k )^2)^{-\frac{3}{2}}$ is the Fourier transform of the
WF of the relative electron-hole motion in the QW exciton: $\phi
(\vec{\rho}) = \sqrt{\frac{2}{\pi \lambda^2}} \cdot
e^{-\frac{\rho}{\lambda}}$; $\lambda$ is associated with the
in-plane Bohr radius of the QW exciton; $S$ is the quantization
area.

In fact, the WF (\ref{WF}) corresponding to the different wave
vectors of the polariton and the electron are nonorthogonal. In
\cite{Malpuech,Ramon} this nonorthogonality was disregarded and
the resulting matrix element was nonunitary:
$M_{\vec{\textbf{K}}_i,\vec{\textbf{k}}_i,\vec{\textbf{Q}},s} \neq
M_{\vec{\textbf{K}}_i+\vec{\textbf{Q}},\vec{\textbf{k}}_i-\vec{\textbf{Q}},-\vec{\textbf{Q}},s}^*$.
In our model, we simply discard the terms in the WF, responsible
for the nonorthogonality of the WF. The matrix element is
calculated as:
\begin{multline}
\label{M}%
M_{\vec{\textbf{K}}_i,\vec{\textbf{k}}_i,\vec{\textbf{Q}},s}=X_{K_i}
X_{|\vec{\textbf{K}}_i+\vec{\textbf{Q}}|} \cdot%
\langle 0|%
\sum_{\vec{\textbf{k}}_2 \neq \vec{\textbf{k}}_i}%
\phi_{\alpha(\vec{\textbf{K}}_i+\vec{\textbf{Q}})-\vec{\textbf{k}}_2}%
\hat{c}_{\vec{\textbf{k}}_i-\vec{\textbf{Q}}}^+
\hat{d}_{\vec{\textbf{K}}_i+\vec{\textbf{Q}}-\vec{\textbf{k}}_2}^+%
\hat{c}_{\vec{\textbf{k}}_2}^+%
\\
\times (\hat{V}_{ee}+\hat{V}_{eh} )%
\sum_{\vec{\textbf{k}}_1 \neq \vec{\textbf{k}}_i-\vec{\textbf{Q}}}%
\phi_{\alpha \vec{\textbf{K}}_i-\vec{\textbf{k}}_1}%
\hat{c}_{\vec{\textbf{k}}_1}^+
\hat{d}_{\vec{\textbf{K}}_i-\vec{\textbf{k}}_1}^+%
\hat{c}_{\vec{\textbf{k}}_i}^+%
|0\rangle.
\end{multline}
The discarding of sum components with $\vec{\textbf{k}}_2 =
\vec{\textbf{k}}_i$ and $\vec{\textbf{k}}_1 =
\vec{\textbf{k}}_i-\vec{\textbf{Q}}$ removes the above mentioned
nonorthogonality of the WF \cite{Nonorthogonality}. The
electron-electron and electron-hole Coulomb interaction operators
$\hat{V}_{ee},\hat{V}_{eh}$ are defined by the following
expressions:
\begin{eqnarray}
\hat{V}_{ee} &=& \frac{1}{2}
\sum_{\vec{\textbf{k}}_1,\vec{\textbf{k}}_2,\vec{\textbf{q}}} V(q)
\cdot \hat{c}_{\vec{\textbf{k}}_1+\vec{\textbf{q}}}^+%
\hat{c}_{\vec{\textbf{k}}_2-\vec{\textbf{q}}}^+%
\hat{c}_{\vec{\textbf{k}}_2}%
\hat{c}_{\vec{\textbf{k}}_1}%
\\
\hat{V}_{eh} &=& -
\sum_{\vec{\textbf{k}}_1,\vec{\textbf{k}}_2,\vec{\textbf{q}}} V(q)
\cdot \hat{c}_{\vec{\textbf{k}}_1+\vec{\textbf{q}}}^+%
\hat{d}_{\vec{\textbf{k}}_2-\vec{\textbf{q}}}^+%
\hat{d}_{\vec{\textbf{k}}_2}%
\hat{c}_{\vec{\textbf{k}}_1},%
\\
V(q) &=& \frac{4 \pi e^2}{\varepsilon S L^2} \cdot%
\frac{e^{-Lq}-1+\frac{5(L q)^3}{8 \pi^2}+\frac{3(L q)^5}{23 \pi^4}}%
{q^3 [1+(\frac{L q}{2 \pi})^2]^2},
\end{eqnarray}
where $V(q)$ is the Fourier transform of the Coulomb potential,
which takes into account the finite QW width $L$ (the electron and
hole transverse WF: $\sqrt{\frac{2}{L}} sin (\pi \frac{z}{L}),
0<z<L$, QW is assumed to be infinitely deep); $e$ is the
elementary charge; $\varepsilon$ is the dielectric susceptibility
of the QW. For the details of calculating $V(q)$ one can refer to
\cite{Ramon}.

Using the anticommutation relation for the fermion operators one
can find from Eq.(\ref{M}) the scattering matrix element:
\begin{eqnarray}
\label{MdMe}
M_{\vec{\textbf{K}}_i,\vec{\textbf{k}}_i,\vec{\textbf{Q}},\pm} = %
X_{K_i} X_{|\vec{\textbf{K}}_i+\vec{\textbf{Q}}|} \cdot %
(M_{\vec{\textbf{K}}_i,\vec{\textbf{k}}_i,\vec{\textbf{Q}}}^{dir} \pm %
M_{\vec{\textbf{K}}_i,\vec{\textbf{k}}_i,\vec{\textbf{Q}}}^{exch}),
\\
\label{Md}
M_{\vec{\textbf{K}}_i,\vec{\textbf{k}}_i,\vec{\textbf{Q}}}^{dir} = %
V(Q) \cdot %
[\frac{1}{(1+(\frac{\beta}{2} \lambda Q)^2)^{\frac{3}{2}}}-%
\frac{1}{(1+(\frac{\alpha}{2} \lambda Q)^2)^{\frac{3}{2}}}],
\\
\label{Me}
M_{\vec{\textbf{K}}_i,\vec{\textbf{k}}_i,\vec{\textbf{Q}}}^{exch} = %
- \frac{2 \lambda^2}{\pi}%
\int \frac{V(k) d^2 k }{%
 (1+\lambda^2 (\vec{\textbf{k}}+\alpha \vec{\textbf{K}}_i - \vec{\textbf{k}}_i+\vec{\textbf{Q}})^2)^{\frac{3}{2}} %
 (1+\lambda^2 (\vec{\textbf{k}}+\alpha \vec{\textbf{K}}_i - \vec{\textbf{k}}_i+\alpha \vec{\textbf{Q}})^2)^{\frac{3}{2}} %
},
\end{eqnarray}
here $\beta=1-\alpha$ is the fraction of the effective mass of a
hole in that of an exciton. One can easily check up that the
obtained matrix element is unitary. Since the matrix element is
calculated, one can calculate the linewidth using
Eq.(\ref{linew}).

\section{Polariton-electron scattering in magnetic field}

We use the similar formalism, as in the previous section, to
calculate the polariton linewidth in a magnetic field
perpendicular to the MC plane ($z$ axis). The field is assumed to
be small and
\begin{equation}
\label{smallB}%
a_H \gg \lambda,
\end{equation}
where $a_H=\sqrt{\frac{\hbar c}{e B}}$ is the magnetic length, $B$
is the magnetic field magnitude, $c$ is the velocity of light.

Magnetic field makes discrete the energy spectrum of free
electrons. We use here the Landau gauge for the vector potential
$\vec{\textbf{A}} = (-By,0,0)$. The electron, hole and exciton WF
are written as \cite{Landau}:
\begin{eqnarray}
\langle \vec{\textbf{r}} | \hat{\tilde{c}}_{n, k_x}^+ |0\rangle &=& \frac{1}{\sqrt{a_H L_x}}%
\cdot e^{i k_x r_x} \cdot%
\varphi_n (\frac{r_y-a_H^2 k_x}{a_H})
\\
\langle \vec{\textbf{r}} | \hat{\tilde{d}}_{n, k_x}^+ |0\rangle &=& \frac{1}{\sqrt{a_H L_x}}%
\cdot e^{i k_x r_x} \cdot%
\varphi_n (\frac{r_y+a_H^2 k_x}{a_H})
\\
\langle \vec{\textbf{R}},\vec{\rho}| \hat{\tilde{b}}_{\vec{\textbf{K}}}^+ | 0 \rangle &=& \frac{1}{S} \cdot%
e^{i K_x R_x+i (K_y+\frac{\rho_x}{a_H^2}) R_y}
\tilde{\phi}(\vec{\rho}),
\end{eqnarray}
where $\varphi_n (x) = \frac{1}{\sqrt{\sqrt{\pi} 2^n n! }} \cdot
e^{-\frac{x^2}{2}} H_n (x)$ is the one-dimensional harmonic
oscillator WF, $H_n(x)$ are the Hermite polynomials; $L_x$ is the
quantization length in the $x$ direction; $\vec{\textbf{R}}$ is
the coordinate of the center of mass of the exciton; $\vec{\rho}$
is the coordinate of the electron-hole relative motion in the
exciton; $\tilde{\phi}(\vec{\rho})$ is the relative motion WF in a
magnetic field. Thus we characterize the free electron motion by
the Landau level number $n$ and by the $x$-component of the
pseudomomentum $k_x$, exciton motion by the total pseudomomentum
$\vec{\textbf{K}}$ \cite{Johnson}. Taking into account
(\ref{smallB}) we can neglect the influence of the magnetic field
on the electron-hole relative motion in an exciton and assume
$\tilde{\phi} (\vec{\rho}) = \sqrt{\frac{2}{\pi \lambda^2}} \cdot
e^{-\frac{\rho}{\lambda}}$. Hence we neglect the change of the
Raby splitting, the polariton dispersion law and the Hopfield
coefficients due to the magnetic field. Precision of such an
approximation is of order of $\frac{\lambda^2}{a_H^2}$
\cite{Johnson}.

Eq. (\ref{linew}) now reads as
\begin{equation}
\label{linewB} %
\gamma_{\vec{\textbf{K}}_i}=4 \pi \sum_{n_i,n_f,k_{i x},\vec{\textbf{Q}},s}%
p_s \cdot %
|\tilde{M}_{\vec {\textbf{K}}_i,n_i,n_f,\vec{\textbf{Q}},s}|^2
F_{n_i} \cdot (1-F_{n_f}) \cdot \delta(E_f-E_i-\hbar \omega_c
(n_f-n_i)),
\end{equation}
here the summation is done not only over the initial and final
electron states and the spin orientation but also over the
$y$-component of the final polariton pseudomomentum $K_{f y}$
since it is not fixed by the conservation law. In Eq.
(\ref{linewB}) $\omega_c = \frac{e B}{m_e c}$ is the electron
cyclotron frequency; $F_n=1/(e^{\frac{\hbar \omega_c
(n+\frac{1}{2})-\mu}{T}}+1)$, Zeeman splitting of the electron and
polariton energy levels is neglected and $p_s$ in Eq.
(\ref{linewB}) corresponds to the isotropic spin distribution. As
a result of the translational symmetry, the absolute value of the
matrix element is not dependent on the $x$-component of the
initial pseudomomentum of the electron $k_{i x}$, and the
summation over $k_{i x}$ gives the constant factor $\frac{S}{2 \pi
a_H^2}$, which is a number of states corresponding to the one
Landau level.

As in the previous section, we use the following WF to calculate
the polariton-electron interaction matrix element:
\begin{equation}
\label{WFB} |\vec{\textbf{K}},n,k_x\rangle=(C_K
\hat{a}_{\vec{\textbf{K}}}^+ + X_K
\hat{\tilde{b}}_{\vec{\textbf{K}}}^+)\cdot
\hat{\tilde{c}}_{n,k_x}^+ |0 \rangle
\end{equation}
(the electron and exciton creation operators in (\ref{WF}) are
replaced by the corresponding operators in a magnetic field). The
WF nonorthogonality is treated as in the previous section. After
the some lengthly calculations, taking into account
(\ref{smallB}), one can find the scattering matrix element in a
magnetic field:
\begin{equation}
\tilde{M}_{\vec{\textbf{K}}_i,n_i,n_f,\vec{\textbf{Q}},s} =%
\frac{a_H^2}{\sqrt{2 \pi}}%
\int \varphi_{n_i}(a_H k_y) \varphi_{n_f}(a_H (k_x-Q_x)) %
e^{- i a_H^2 k_x (k_y-Q_y)} \cdot%
M_{\vec{\textbf{K}}_i, \vec{\textbf{k}}, \vec{\textbf{Q}}, s} d^2 k, %
\end{equation}
where $M_{\vec{\textbf{K}}_i, \vec{\textbf{k}}, \vec{\textbf{Q}},
s}$ is calculated using Eq.(\ref{MdMe})-(\ref{Me}). $k_{i x}$
arises only in the phase factor, which does not contribute to the
scattering probability and is omitted here.

Precision of the above calculations is of order of
$\frac{\lambda^2}{a_H^2}$.

\section{Results and discussion}

The numerical calculations was performed for the GaAs based MC
with the QW of width $L = 75 \AA$ and with Raby splitting $4$ meV.
The exciton Bohr radius $\lambda = 150 \AA$, the ratio of electron
and exciton effective masses $\alpha=0.17$, the electron gas
concentration in the QW $n_e = 10^{9}$ cm$^{-2}$.

\begin{figure}
\includegraphics{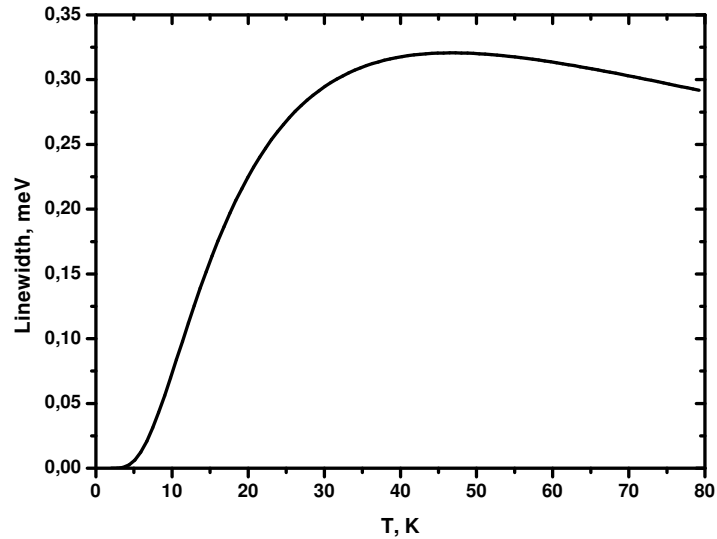} %
\caption{Dependence of the polariton linewidth on the electron
gas temperature for zero exciton-photon detuning and $B=0$.} %
\label{WT}
\end{figure}

Fig.\ref{WT} shows the calculated temperature dependence of the
polariton linewidth for zero exciton-photon detuning
($E_{ph}(k=0)-E_{ex}(k=0)=0$), corresponding to $\Delta E = 2$
meV, without magnetic field. For small temperatures ($T\ll \Delta
E$) the polariton-electron scattering is noneffective since
polaritons can not receive from electrons sufficient energy to get
to the reservoir with high $k$ (Fig.\ref{dispersion}a). The
polariton linewidth increases with increasing temperature and
after reaching a maximum slowly decreases. Such a decrease for
high temperatures is due to the smaller exchange scattering matrix
element for high energy electrons (see Eq. (\ref{Me})).

\begin{figure}
\includegraphics{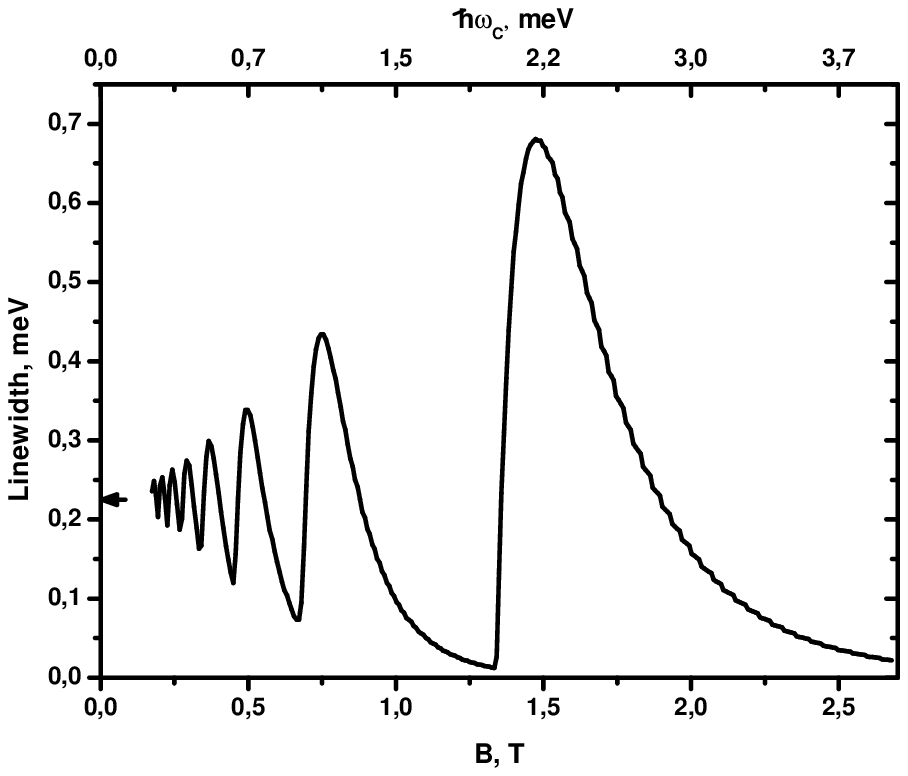} %
\caption{Dependence of the polariton linewidth on the magnetic
field magnitude for zero exciton-photon detuning and $T=20$. The
top scale shows the separation between Landau levels. Arrow
indicates the value of the polariton linewidth, calculated for
$B=0$ and $T=20K$
(see fig.\ref{WT}).} %
\label{WB}
\end{figure}

Fig.\ref{WB} shows the calculated magnetic field dependence of the
polariton linewidth for zero exciton-photon detuning and $T=20$K.
Maximums in Fig.\ref{WB} correspond to $\hbar \omega_c
(n_i-n_f)\gtrapprox \Delta E$ (see the top scale in the figure and
note that $\Delta E = 2$ meV). First maximum corresponds to
$n_i-n_f=1$ (Fig.\ref{dispersion}b), second maximum corresponds to
$n_i-n_f=2$ (Fig.\ref{dispersion}c) etc. Thus polariton-electron
scattering rate can be not only decreased, but also increased by a
magnetic field perpendicular to the MC plane. For the magnetic
fields, for which $\hbar \omega_c$ considerably exceeds the LP
branch depth (Fig.\ref{dispersion}d), the polariton linewidth
monotonically decreases with the increasing field magnitude
(Fig.\ref{WB}). The last effect was observed experimentally for
$B=7$ T \cite{Qarry}. Unfortunately there is no experimental data
concerning the MC polariton-electron scattering in weak magnetic
fields known.

Similar oscillatory magnetic field dependence of scattering rate
was derived theoretically for electron-electron scattering between
two QW subbands \cite{Kempa}. Although the difference of the
considered systems here and in \cite{Kempa}, the nature of the
scattering rate oscillations is the same.

\acknowledgments

The author wishes to thank N.N. Sibeldin, M.L. Skorikov and V.A.
Tsvetkov for fruitful discussions, V.G. Bilykh and N.A. Gippius
for careful reading and critical remarks and D.B. Volegov for help
in preparation of the manuscript. The work was supported by RFBR
(project 05-02-17328), INTAS (project 01-0832), basic research
program of RAS presidium ``Low dimensional quantum structures''
(project project 8.7) and leading scientific schools state support
program (project 1923.2003.2).


\begin{thebibliography}{99}
\bibitem{Nonorthogonality}
In the limit $\alpha \rightarrow 0$, the orthogonal WF could be
chosed naturally, if one would use the WF of an electron,
propagating in the field of an infinitly heavy hole instead of the
WF of a free electron. One can proove directly, that the matrix
element derived by such a way coinsides with the matrix element
calculated using the WF nonorthogonality removal in the limit
$\alpha \rightarrow 0$.
\end{thebibliography}
\end{document}